\documentclass[twocolumn,10pt]{aa}
\usepackage{graphicx}
\usepackage{txfonts}
\begin{document}
   \title{On the kinematics of the neutron star\\
    low mass X-ray binary \mbox{Cen X-4}}

   \subtitle{}

   \author{J. I. Gonz\'alez Hern\'andez\inst{1},  R. Rebolo\inst{1,2,3}, J.
   Pe\~narrubia\inst{3}, J. Casares\inst{1},\and G. Israelian\inst{1}}

   \offprints{J. I. Gonz\'alez Hern\'andez}

   \institute{Instituto de Astrof{\'\i }sica de Canarias, E-38205 La Laguna,
   Tenerife, Spain \\ \email{}jonay@iac.es, rrl@iac.es, jcv@iac.es,
   gil@iac.es 
         \and
   Consejo Superior de Investigaciones Cient{\'\i }ficas, Spain
         \and
   Max-Planck Institut f\"ur Astronomie, K\"onigstuhl, 17, D-69117 Heidelberg,
   Germany \\ \email{jorpega@mpia-hd.mpg.de}
             }

   \date{}

\abstract{
We present the first determination of the proper motion of the neutron star
low mass X-ray binary \mbox{Cen X-4} measured from relative astrometry of the 
secondary star using optical images at different epochs. We determine the
Galactic space velocity components of the system and find them to be
significantly different from the mean values that characterize the kinematics
of stars belonging to the halo, and the thin and the thick disc of the
Galaxy. The high metallicity of the secondary star of the system rules out a
halo origin and indicates that the system probably originated in the Galactic
disc. A statistical analysis of the galactocentric motion revealed that this
binary moves in a highly eccentric ($e\simeq 0.85\pm0.1$) orbit with an
inclination of $\simeq 110^\circ$ to the Galactic plane. The large Galactic
space velocity components strongly support that a high natal kick as a result
of a supernova explosion could have propelled the system into such an orbit
from a birth place in the Galactic disc. The high Li abundance in the
secondary, comparable to that of stars in star forming regions and young
stellar clusters like the Pleiades, may suggest a relatively recent
formation of the system. Following the orbit backwards in time, we found that
the system could have been in the inner regions of the Galactic disc
$\sim$100--200 Myr ago. The neutron star might have formed at that moment.
However, we cannot rule out the possibility that the system formed at a much
earlier time if a Li production mechanism exists in this LMXB.

}

\authorrunning{Gonz\'alez Hern\'andez et al.}
\titlerunning{On the kinematics of the LMXB \mbox{Cen X-4}}

\maketitle
%
%________________________________________________________________

\section{Introduction}

\mbox{Cen X-4} (V822 Cen) was discovered by the {\it Vela 5B} satellite
during an X-ray outburst in 1969 (Conner et al.\ 1969). During the decay of a
second outburst in 1979, a type I X-ray burst was observed, which indicated
that the X-ray source is a neutron star (Matsuoka et al.\ 1980). This system
is a prototype low mass X-ray binary (LMXB) consisting of a compact object of
$M_{\rm NS} =$ 0.5--2.1 {$M_\odot$} (Shahbaz et al.\ 1993) and a donor star
with a mass in the range 0.04 {$M_\odot$} $< M_2 < 0.58$ {$M_\odot$} (Torres 
et al.\ 2002).  

The high Galactic latitude ($b\sim 24^\circ$) of the system coupled with its
distance ($\sim 1.2\pm0.3$ kpc, Chevalier et al.\ 1989) places the object at
the Galactic height of $\sim$0.4--0.6 kpc and might suggest that it belongs
to the Galactic halo. However, its relatively large centre-of-mass radial
velocity ($\gamma =191 \pm 1$ {${\rm km}\:{\rm s}^{-1}$}, Casares et al.\ 2005, in
preparation) opens the possibility that the system received a \emph{natal}
kick as a result of the supernova explosion that generated the compact
object and it might come from the Galactic disc. Such kicks have been
proposed to explain the large transverse motions of neutron stars on the
plane of the sky (e.g.\ Lyne \& Lorimer 1994).   

The kinematic properties of other similar systems at high Galactic latitude
have been studied by Mirabel et al.\ (2001, \mbox{XTE J1118+40}) and Mirabel
\& Rodr{\'\i}gues (2003, \mbox{Sco X-1}) by measuring their proper motions. 
These authors argued that \mbox{XTE J1118+40} and \mbox{Sco X-1} have
kinematics consistent with the most ancient stars and globular clusters in
the inner Galactic halo. However, according to the the galactocentric orbits
of these objects, an origin in the bulge or disc of the Galaxy cannot be
ruled out. In fact, Gualandris et al.\ (2004) have recently suggested that
the black hole binary \mbox{XTE J1118+40} could have formed in the Galactic
disc $\sim$2--5 Gyr ago. 
   
The kinematic properties of \mbox{Cen X-4} can help us to understand its
origin. Here, we report the first determination of the proper motion of this
system via relative astrometry of the secondary star using optical images.
This, together with the radial velocity and the distance, allows us to
determine the kinematic parameters and study the galactocentric orbit of this
important LMXB. 
   
%__________________________________________________________________

\section{Proper motion}

Accurate proper motion measurements of LMXBs have been obtained
using very long baseline array (VLBA) or very long baseline interferometry
(VLBI) techniques at radio wavelengths and/or from optical observations of
these systems at HST or large telescopes. Unfortunately, there are no 
reported observations of \mbox{Cen X-4} with either of these facilities.
However, several campaigns have carried out observations of this target using
optical ground-based telescopes. We selected two images among those with
better pixel scales: a plate from the Second Epoch Survey\footnote{Plates
from this survey, also called Second Palomar Sky Survey (POSSII), have been
digitized and compressed by the ST ScI. The digitized images are copyright
(c) 1993--2000 by the Anglo--Australian Observatory Board and are distributed
herein by agreement. All Rights Reserved. Produced under Contract No.
NAS5-2555 with the National Aeronautics and Space Administration.} of the 
Southern Sky observed in May 1994 in the Anglo--Australian Observatory (AAO)
with the UK Schmidt Telescope with a pixel scale of 1 arcsec, and a second
image with a pixel scale of 0.238 arcsec obtained in May 2002 with the Wide
Field Imager at the European Southern Observatory (ESO), Observatorio La
Silla, using the Cassegrain focus of the 2.2 m MPG/ESO telescope. 

\begin{figure}[ht!]
  \centering
  \includegraphics[width=7.5cm,angle=0]{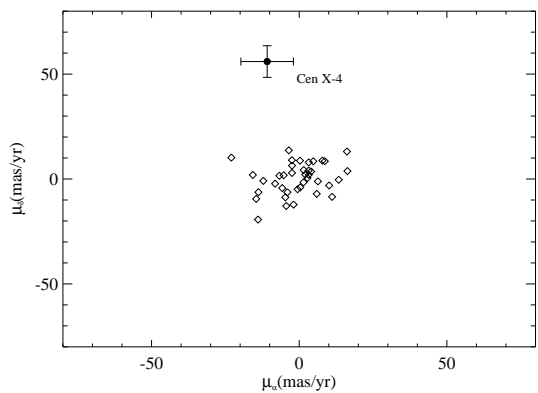}
  \caption{Distribution of shifts in right ascension and declination for the
  39 reference stars and the target (wide cross), \mbox{Cen X-4}. The error
  bars give the total error (see \S 2).\label{fig1}} 
\end{figure}

The proper motion was determined using several tasks within IRAF. First, we
measured the pixel coordinates of 40 stars including the target, \mbox{Cen
X-4}, in both images (with the task {\scshape imcentroid}). Second, the 1994
image was registered on the 2002 one, by applying the coordinate
transformation (using the tasks {\scshape geomap} and {\scshape geoxytrans})
between both frames of reference objects. We selected as reference objects
those stars whose transformation residuals were lower than 2$\sigma$ above
the rms of the transformation (first criterion). Third, we measured the
relative shifts of all the stars between their positions corresponding to the
1994 and 2002 images in the reference frame of the 2002 image. We then
rejected those stars whose proper motions were higher than 3$\sigma$ above
the standard deviation of the motions of all the stars (second criterion).
The whole procedure was iterated until we could match the reference objects
with both criteria unambiguously. We aligned (using the tasks {\scshape
ccmap} and {\scshape cctran}) the 2002 frame in right ascension and
declination with these reference objects using the coordinates given by 
2MASS\footnote{The Two Micron All Sky Survey is a joint project of the
University of Massachusetts and the Infrared Processing and Analysis
Center/California Institute of Technology, funded by the National Aeronautics
and Space Administration and the National Science Foundation.}. In order to
correct for geometrical distorsions with {\scshape geomap} and {\scshape
ccmap}, we selected general geometry transformations up to order 2. These take into account rotation, 
pixel/celestial coordinate scale and shift between the reference frames of
both images. Finally, we computed the relative displacements of our target
and estimated the error from the average relative motions of the reference
stars (see Fig.\ 1). The computed proper motion is $\mu_\alpha =  11 \pm 10$
mas/yr and $\mu_\delta = -56 \pm  10$ mas/yr, which leads to an overall
yearly motion of $\mu = -57 \pm  12$ mas/yr along a position angle with
respect to the North Celestial Pole of  $100^\circ \pm  12^\circ$. At a
distance $d$ (in kpc), this proper motion corresponds to a transverse
velocity on the plane of the sky of $(270 \pm 56)d$ {${\rm km}\:{\rm
s}^{-1}$}.  

\section{Distance to \mbox{Cen X-4}}

In order to determine the space velocity of the system from the proper motion
measurement we need to know its distance to the Sun. The optical properties
of the secondary star can provide an estimate of the distance, although we
found several discrepancies among the values obtained using different
photometric filters. The mass range estimated for the secondary star of
$0.04<M_2/M_\odot<0.58$ (Torres et al.\ 2002) leads to a stellar radius of
$0.5<R_2/R_\odot<1.2$ by assuming that the secondary fills its Roche lobe.
This radius, together with our spectroscopic estimate of the effective
temperature, $T_{\rm eff}=4500\pm100$ K (Gonz\'alez Hern\'andez et al.\
2005), provides an intrinsic bolometric luminosity of
$0.095<L_2/L_\odot<0.371$. The apparent magnitude is $m_{V}\sim18.2$ (van
Paradijs et al.\ 1980) and the colour excess, $E_{B-V}=0.1\pm0.05$ (Blair et
al.\ 1984). Using the bolometric corrections (Bessel et al.\ 1998) and the
effective temperature, surface gravity and veiling from the accretion disc
derived by Gonz\'alez Hern\'andez et al.\ (2005, submitted), we found
$1.5<d_{m_{V}}({\rm kpc})<3.6$. Note that \mbox{Cen X-4} is highly variable
in quiescence, with reported magnitudes $m_{\rm V}\sim$ 18--18.7 (Chevalier
et al.\ 1989; Cowley et al.\ 1988). From the $m_{R}\sim17.5$ and
$m_{H}\sim15.1$ (Shahbaz et al.\ 1993) and assuming $A_{R}/A_{V}\sim0.83$ and
$A_{H}/A_{V}\sim0.21$ (Schaifers et al.\ 1982), we obtained
$1.2<d_{m_{R}}({\rm kpc})<2.9$ and $0.9<d_{m_{H}}({\rm kpc})<2.3$. The
distance obtained from the $V$ and $R$ filters in particular are rather
uncertain because our veiling estimates and the assumed apparent magnitudes
were determined at different epochs and are also affected by the contribution
of a contaminant star (with $m_{V}\lesssim 20.5$) to the total flux (Casares
et al.\ 2005, in preparation). 

\begin{figure*}[ht!]
  \centering
  \includegraphics[width=11cm,angle=0]{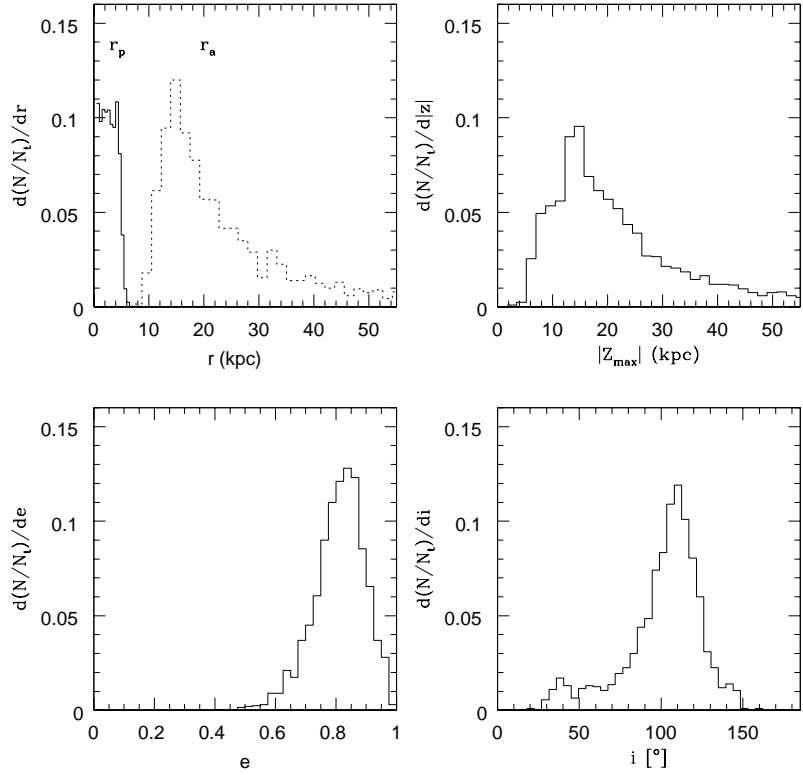}
  \caption{Distribution of perigalacticon and apogalacticon distances
  (upper left panel), maximum vertical distance (upper right panel),
  eccentricities (lower left panel) and orbital inclinations (lower right
  panel). Distributions are normalized to the total number of orbits in our
  calculus ($N_t=$ 5000).\label{fig2}}
\end{figure*}

A more confident value of the distance can be calculated from the peak fluxes
of the X-ray precursor observed in the 1969 outburst (Belian et al.\ 1972)
and the X-ray burst detected during the 1979 outburst (Matsuoka et al.\
1980). From the observed fluxes of $\sim1.4\times10^{-6}$ erg cm$^{-2}$
s$^{-1}$ and $\sim1\times10^{-6}$ erg cm$^{-2}$ s$^{-1}$ (for 1969 and 1979,
respectively) and by assuming that they correspond to X-ray luminosities at
the Eddington limit for a canonical neutron star ($L_{\rm Edd} \sim$
1.6$-$2.7 $\times 10^{38}$ erg s$^{-1}$), Chevalier et al.\ (1989) derived a
distance of $1.2\pm0.3$ kpc. This measurement can also be estimated by
considering recent average peak luminosities of photospheric radius expansion
bursts observed in globular clusters, which is (2--3.8) $\times 10^{38}$ erg
s$^{-1}$ (Jonker \& Nelemans 2004, and references therein). From these peak
luminosities and the observed fluxes we obtained a distance of
$d\simeq1.4\pm0.3$ kpc.   

The space velocity calculated according to the derived proper motion exceeds
the Galactic escape velocity (around 500 {${\rm km}\:{\rm s}^{-1}$} in the
solar neighbourhood) if the system is further than 1.7 kpc from the Sun (see
\S 4). Therefore, we  assume hereafter a distance to \mbox{Cen X-4} of
$0.9<d({\rm kpc})<1.7$ which is where both the X-ray and optical distance
measurements overlap.       

\section{Space velocity}

From the values of position, distance, proper motion and radial velocity, the
velocity components $U_{\rm LSR}$, $V_{\rm LSR}$ and $W_{\rm LSR}$ relative
to the Local Standard of Rest can be calculated using, for example, Johnson
\& Soderblom's (1987) equations of transformation. Assuming that the Sun
moves with components $(U_{\odot,{\rm LSR}},V_{\odot,{\rm LSR}},W_{\odot,{\rm
LSR}})=(10.,5.2,7.2)$ {${\rm km}\:{\rm s}^{-1}$} (Dehnen \& Binney 1998)
relative to the Local Standard of Rest, from the two extreme values of the
distance we obtained: for $d=0.9$ kpc, $(U_{\rm LSR},V_{\rm LSR},W_{\rm LSR})
= (204\pm21,-196\pm35,-122\pm38)$ {${\rm km}\:{\rm s}^{-1}$}, and for $d=1.7$
kpc, $(U_{\rm LSR},V_{\rm LSR},W_{\rm LSR}) = (239\pm40,-303\pm67,-305\pm71)$
{${\rm km}\:{\rm s}^{-1}$}. This implies that the source moves towards the
Galactic Centre and has a 3$\sigma$ vertical motion towards the Galactic
plane. These two sets of values are rather different from the mean values of
$U$, $V$ and $W$ that caracterize the kinematics of stars  belonging to the
halo, and the thin and thick discs of the Galaxy (Chiba \& Beers 2000;
Soubiran et al. 2003). The system velocities would be $v_{\rm sys}=308\pm51$
{${\rm km}\:{\rm s}^{-1}$} and $v_{\rm sys}=493\pm105$ {${\rm km}\:{\rm
s}^{-1}$} for $d=0.9$ kpc and $d=1.7$ kpc respectively. 

\begin{figure*}[ht!]
  \centering
  \includegraphics[width=14cm,angle=0]{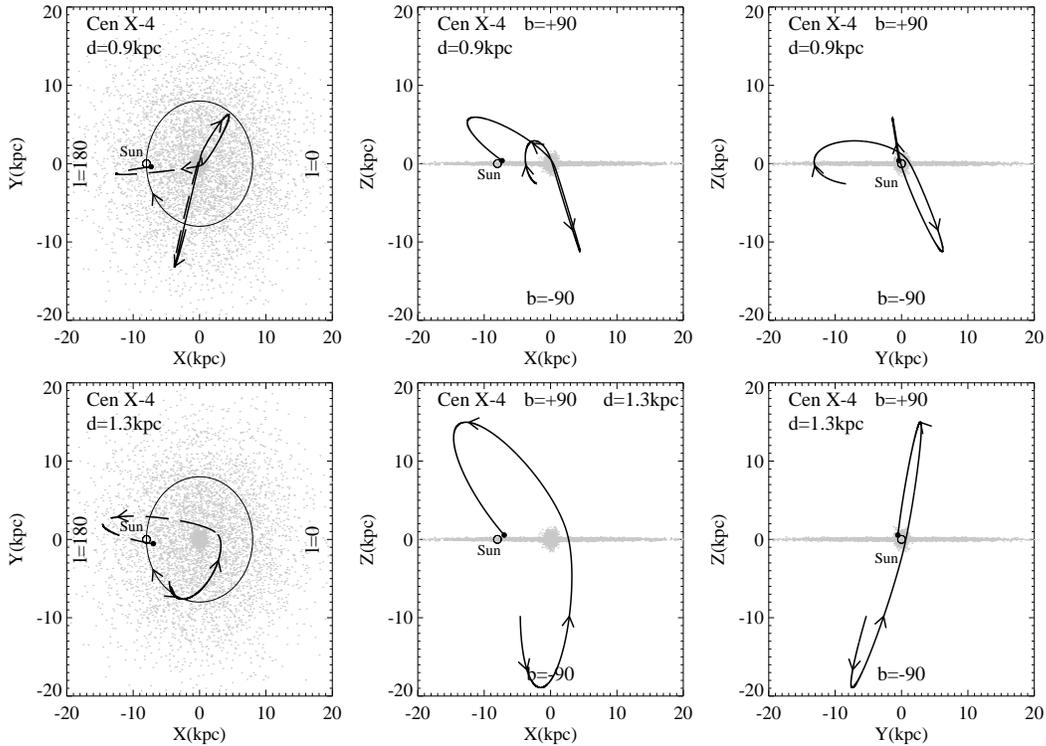}
  \caption{Galactocentric orbits of \mbox{Cen X-4} viewed from above the Galactic
  plane (left panels) and edge on (centre and right panels). The upper
  panels show the orbit for a heliocentric distance of $d=0.9$ kpc and the
  lower panels for $d=1.3$ kpc. In the left panels, the solid line
  represents the section of the orbit in the Southern Galactic Hemisphere,
  whereas the dashed line shows the section of the orbit in the Northern
  Galactic Hemisphere. Both orbits were calculated by integrating
  backwards over 460 Myr. The circular orbit corresponds to the
  Sun. Open and filled circles show the current positions of the Sun
  and \mbox{Cen X-4} respectively.\label{fig3}} 
\end{figure*}

\section{Galactocentric orbit}

We integrated the orbit of \mbox{Cen X-4} based on the heliocentric position
and kinematic data listed in Table 3. In order to convert heliocentric
positions and velocities into coordinates in the Galaxy frame we placed the
Sun at $(X_\odot,Y_\odot,Z_\odot)=(-8,0,0)$ kpc, with a velocity $\dot {\bf
r}_\odot\equiv (U_\odot,V_\odot,W_\odot)=(10.0 ,225.2 ,7.2)$ (Binney \&
Merrifield 1998; Dehnen \& Binney 1998) and used the equation   
\begin{eqnarray}
\dot{\bf r}=\dot{\bf r}_\odot + v_{\rm r}(\cos b \cos l,\cos b\sin l,\sin
b)\nonumber\\+r_s\mu_l \kappa(-\sin l,\cos l,0)\nonumber\\
+r_s\mu_b \kappa(-\sin b \cos l,-\sin b \sin l, \cos b)\nonumber
\end{eqnarray}
where $r_s$ is the heliocentric distance, $\kappa\simeq 4.74$, a conversion
factor from (kpc mas/yr) to ({${\rm km}\:{\rm s}^{-1}$}), $v_r$, the
heliocentric radial velocity and $\mu_l,\mu_b$, proper motions in the
longitudinal and latitudinal direction, respectively. 

We assumed a three-component Galaxy model that consists of: (i) a Miyamoto \&
Nagai (1975) disc, (ii) a Hernquist bulge (Hernquist 1990) and (iii) a
spherical logarithmic halo. The Galaxy parameters can be found, for example,
in Law et al. (2004). 

Unfortunately, the heliocentric distance and velocity of \mbox{Cen X-4} are
poorly determined. Values with these large errors represent a large set
of possible orbits that must be statistically treated. We calculated an
initial set of 5000 orbits using Monte Carlo simulations, where the
heliocentric distance is randomly chosen within the range given above. Proper
motions were distributed following a Gaussian with variance equal to the
measured error of 10 mas/yr and mean value $\langle\mu_l\rangle=-22$ mas/yr
and $\langle \mu_b\rangle=-53$ mas/yr. Subsequently, we integrated each orbit
1 Gyr backwards in time. In Fig.\ 2 we show the orbital parameters that
result from the distribution of initial positions and velocities. The orbit
of \mbox{Cen X-4} appears totally inconsistent with typical orbits of Milky
Way stars. This system has a perigalacticon of $r_p\sim 1-4$ kpc, whereas the
apogalacticon occurs at $r_a\sim 14$--16 kpc from the Galactic Centre.
Although $r_p$ and $r_a$ appear to be fairly sensitive to the initial
position and velocity, the resulting eccentricity, $e\equiv
(r_a-r_p)/(r_a+r_p)=0.85\pm 0.10$, barely depends on the errors from
heliocentric distance and proper motion  measurements. This eccentricity is
considerably higher than that of disc stars ($e\simeq 0.5$--0.1). The orbital
inclination from the disc plane also shows a well defined distribution peak
at $i\simeq 110^\circ$ and determines its orbital sense of motion to be
retrograde for most of the distances considered in the distribution. This
orbital inclination is considerably higher than for Milky Way stars (i.e.\
$i_\odot=4^\circ$) and leads to high vertical motions. As we can see in the 
upper right panel, \mbox{Cen X-4} will move (or may have moved) in halo
regions, with a most likely maximum vertical distance as large as 12--16 kpc.
   
In Fig.\ 3 we have represented the galactocentric orbits of \mbox{Cen X-4}
for two distances, 0.9 and 1.3 kpc. These orbits were calculated by
integrating backwards in time for 460 Myr and both orbits are highly
eccentric. The orbit for $d=0.9$ kpc shows a prograde sense of motion,
whereas for $d=1.3$ kpc the system describes a retrograde orbit. These orbits
intersect the disc plane for the first time (integrating backwards in time)
148 and 243 Myr ago for $d=0.9$ kpc and $d=1.3$ kpc, respectively.

\section{Pinning down the formation region of \mbox{Cen X-4}}

\begin{figure*}[ht!]
  \centering
  \includegraphics[width=11cm,angle=0]{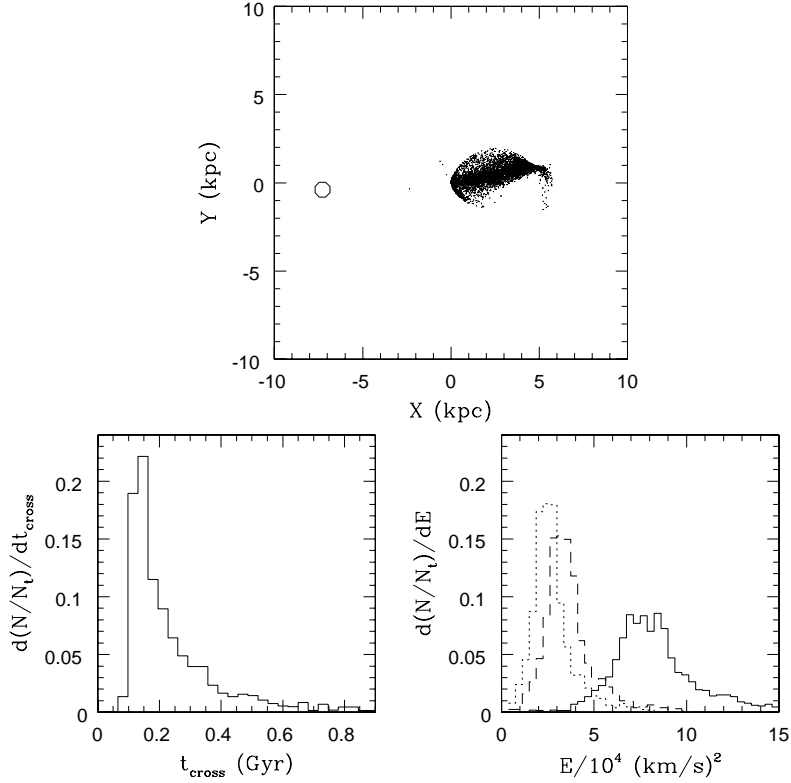}
  \caption{Upper panel: First crossing of the axisymmetry plane when
integrating the orbit of Cen X-4 backwards in time. The circle indicates its
present position, and its diameter corresponds to the uncertainty in the
distance estimate. Lower left panel: Distribution of times when the orbit
(integrated back in time) crosses the disc plane for the first time. Lower right
panel: distribution of energies per unit of mass gained by Cen X-4 in
comparison with circular orbits at the first-crossing position. The solid
line represents the definition for the energy $E$ given in the text whereas
the dotted and dashed lines show $E'$ for $M_{\rm He}=4$ {$M_\odot$} and
$M_2=0.8$ {$M_\odot$} for $M_{\rm He}=3$ {$M_\odot$} and $M_2=0.5$
{$M_\odot$} respectively.\label{fig4}} 
\end{figure*}

Two main scenarios have been proposed to explain the current kinematic
properties of LMXBs observed at high Galactic latitudes: (i) these binary
systems originated in the Galactic bulge or in the disc and received a natal
kick as a result of black hole/neutron star formation, or (ii) they
formed in the Galactic halo either in dense globular clusters or as isolated
ancient primordial binaries. Mirabel \& Rodr{\'\i}guez (2003) argued that the
LMXB \mbox{Sco X-1} most probably was formed in the core of a globular
cluster in a close encounter of the compact object with a single star or with
a binary (see Verbunt 2003 for a review).

Halo stars and globular cluster stars are generally metal-poor (with heavy
elements between 10 and 1000 times less abundant than in the Sun). If the
Cen X-4 system had originated in the halo, the secondary star would have a
low metallicity. Recently, the iron content\footnote{$[{\rm Fe}/{\rm
H}]=\log [N(\mathrm{Fe})/N(\mathrm{H})]_\star -\log 
 [N(\mathrm{Fe})/N(\mathrm{H})]_\odot$.} of the secondary star has been 
determined at roughly twice solar ($[{\rm Fe}/{\rm H}]=0.23\pm0.1$,
Gonz\'alez Hern\'andez et al.\ 2005). This high metal content, together with
measured element abundance ratios in the companion star, makes it very
unlikely that the system was formed in the Galactic halo, either in a
globular cluster or as an ancient primordial binary. Most probably, the
system formed in the Galactic disc and was expelled from its birth place as a
consequence of a supernova (SN hereafter) explosion of the massive progenitor
of the neutron star.  
 
An origin in the Galactic disc requires that the system obtained enough energy
to be expelled, most probably as a consequence of a natal kick in a SN
explosion. In the lower right panel of Fig.\ 4 we show an estimate of the
energy per unit of mass gained by the binary system due to a supernova at a
first crossing. We define this energy as $E=E_{\rm orb}-E_{\rm circ}$, where
$E_{\rm circ}$ is the energy per unit of mass of a star following a circular
orbit in the region where \mbox{Cen X-4} crosses the Galactic plane. The
resulting distribution is close to a Gaussian with $\langle E\rangle \simeq
(8\pm 1)\times 10^4 ({\rm km}\:{\rm s}^{-1})^2$. This energy must be lower
than the difference between the total energy of the system after and before
the SN explosion if the binary components remain bound. Brandt \&
Podsiadlowski (1995) investigated the effects of high SN kick
velocities on the orbital parameters of post-SN neutron star binaries.
Using their formalism we compared the pre-SN, $E_{\rm pre-SN}$, and
post-SN, $E_{\rm post-SN}$, total energies per unit of mass of the system
after undergoing SN explosions.  We defined this energy balance as $E'
=E_{\rm post-SN} (M_{\rm T,post-SN}/M_{\rm T,pre-SN})-E_{\rm pre-SN}$ where
$M_{\rm T,post-SN}$ and $M_{\rm T,pre-SN}$ are the total masses of the system
after and before the SN event respectively. We found that energy released in
a SN of He core progenitor of mass $M_{\rm He}=4$ {$M_\odot$} and a secondary
star of $M_2=0.8$ {$M_\odot$} is $E' \sim 1-3\times 10^4 ({\rm km}\:{\rm
s}^{-1})^2$, whereas for $M_{\rm He}=3$ {$M_\odot$} and $M_2=0.5$
{$M_\odot$}, $E' \sim 1-2\times 10^4 ({\rm km}\:{\rm s}^{-1})^2$. These
energies are consistent with the energy required to launch \mbox{Cen X-4}
into an orbit like this from a birth place in the Galactic disc (see Fig.\
4). A detailed study of the distribution of the kick directions with respect
to the orbital motions of each component shows that it strongly depends on
the masses of the different components of the system before and after the SN
explosion. Thus the kick direction cannot randomly be chosen and is more
restricted when the differences between the masses of both components are
higher. 

The abundance of lithium in the secondary star of \mbox{Cen X-4} ($\log
\epsilon(\mathrm{Li})_{\rm LTE}=3.06 \pm 0.29$) is close to 
the cosmic Li abundance in the Galactic disc (Mart{\'\i}n et al.\ 1994a;
Gonz\'alez Hern\'andez et al.\ 2005) and thus substantially higher than 
in stars of comparable mass of the young Pleiades cluster whose age is $\sim
120$ Myr (Mart{\'\i}n et al.\ 1998, Stauffer, Schultz \& Kirkpatrick 1998). 
This might indicate that the system is relatively young since Li is severely
depleted in the atmospheres of such stars during  pre-main sequence and main
sequence evolution. In order to explore the possibility of a recent formation
of \mbox{Cen X-4} in the Galactic disc, we calculated the positions where the
orbit intersects the Galactic plane for the first time (integrating backwards
in time). The result is plotted in the upper panel of Fig.\ 4, which shows
that the orbit crosses the inner region of the Milky Way in the range 0--4
kpc from the Galactic Centre. The distribution of times (lower left panel)
indicates that the system was probably in the Galactic disc $\sim$100--200
Myr ago. The high Li abundance of the secondary star might be explained if
the system had formed at that moment. However, it has been proposed that a Li
production mechanism via spallation reactions could be responsible for the
high Li abundances of LMXBs (Mart{\'\i}n et al.\ 1994b; Yi \& Narayan 1997;
Guessoum \& Kazanas 1999). No direct evidence for such a mechanism has been
found yet in \mbox{Cen X-4} or in any other LMXB. Nevertheless, we cannot
rule out that such a mechanism has been active in \mbox{Cen X-4} and, thus,
that the system could be much older.

\section{Conclusions}

We have used optical relative astrometry of the secondary star of the neutron
star binary \mbox{Cen X-4} to measure its proper motion. This proper motion,
together with the estimated distance and radial velocity and its celestial
coordinates, allows us to determine the space velocity components. These
appear to be rather different from the mean values of $U$, $V$ and $W$ for
stars belonging to the halo, the thin and the thick disc of the Galaxy. The
system is falling into the Galactic plane with a large vertical velocity
$\gtrsim$ 100 {${\rm km}\:{\rm s}^{-1}$}.   

The high metallicity ($[{\rm Fe}/{\rm H}]=0.23\pm0.1$, Gonz\'alez Hern\'andez
et al. 2005) and the element abundance ratios of the companion star in
\mbox{Cen X-4} makes it very unlikely that the system formed in the Galactic
halo. Most probably, the system originated in the Galactic disc and was
propelled into its current position by a natal kick in a supernova explosion.
   
We have explored the galactocentric orbit of \mbox{Cen X-4} through a
statistical analysis that takes into account the uncertainty in the proper
motion and the distance. The orbital inclination from the Galactic plane is
$\simeq 110^\circ$ and determines its orbital sense of motion to be
retrograde in a highly eccentric ($e\simeq 0.85\pm0.1$) orbit. Following this
orbit backwards, we have found that the system was probably at the 
Galactic disc $\sim$100--200 Myr ago. The formation of the neutron star at
that time would imply a relatively young system and may provide a natural
explanation for the observed high Li abundance of the secondary star.
However, if there is an active  Li production  mechanism in \mbox{Cen X-4},
the Li  abundance cannot be directly interpreted as an indication of youth,
and it cannot be discarded that the system formed at much earlier times. 
 
\begin{acknowledgements}

This work has made use of IRAF facilities and has been partially financed by
the Spanish Ministry projects AYA2001-1657 and AYA2002-03570. 

\end{acknowledgements}

\end{document}